\RequirePackage{fix-cm}
\documentclass[twocolumn,epjc3]{svjour3}      
\RequirePackage{graphicx}
\RequirePackage{mathptmx}      
\RequirePackage{flushend}
\RequirePackage[numbers,sort&compress]{natbib}
\RequirePackage[colorlinks,citecolor=blue,urlcolor=blue,linkcolor=blue]{hyperref}
\RequirePackage[T1]{fontenc}
\journalname{Eur. Phys. J. C}
\smartqed  
\newcommand\beq{\begin{eqnarray}}
\newcommand\eeq{\end{eqnarray}}
\newcommand\bq{\begin{equation}}
\newcommand\eq{\end{equation}}
\newcommand\nutqpet{\mbox{\boldmath $(\hat{\eta}_{\nu})^{ \perp}$}\cdot({\bf \hat{q} \times
(\hat{p}_{e})^\perp})}
\newcommand\netqpet{\mbox{\boldmath $(\hat{\eta}_{e})^{\perp}$}\cdot({\bf \hat{q} \times (\hat{p}_{e})^{\perp}})}
\newcommand\petnetnut{{\bf (\hat{p}_{e})^{\perp}}\cdot( \mbox{\boldmath $(\hat{\eta}_{e})^{\perp}$}\times \mbox{\boldmath $(\hat{\eta}_{\nu})^{ \perp}$})}
\newcommand\qnetnut{{\bf \hat{q}}\cdot( \mbox{\boldmath $(\hat{\eta}_{e})^{\perp}$}\times \mbox{\boldmath $(\hat{\eta}_{\nu})^{ \perp}$})}
\newcommand\netnut{\mbox{\boldmath $(\hat{\eta}_{e})^{\perp}$}\cdot\mbox{\boldmath $(\hat{\eta}_{\nu})^{ \perp}$}}
\newcommand\nutpet{\mbox{\boldmath $(\hat{\eta}_{\nu})^{ \perp}$}\cdot{\bf (\hat{p}_{e})^{\perp}}}

\newcommand\netpet{\mbox{\boldmath $(\hat{\eta}_{e})^{\perp}$}\cdot{\bf (\hat{p}_{e})^{\perp}}}
\DeclareSymbolFont{matha}{OML}{txmi}{m}{it}
\DeclareMathSymbol{\vv}{\mathord}{matha}{29} 

\begin{document}
\title{On possibility of time reversal symmetry violation in neutrino elastic scattering on polarized electron target} 
\author{W. Sobk\'ow\thanksref{e1,addr1} \and A. B\l{}aut\thanksref{e2,addr1} 
}
\thankstext{e1}{e-mail: wieslaw.sobkow@ift.uni.wroc.pl}
\thankstext{e2}{e-mail: arkadiusz.blaut@ift.uni.wroc.pl}

\institute{ Institute of Theoretical Physics, University of Wroc\l{}aw,
Pl. M. Born 9, PL-50-204~Wroc{\l}aw, Poland\label{addr1}
}

\date{Received: date / Accepted: date}

\maketitle
\begin{abstract}

In this paper, we indicate a possibility of utilizing  the  elastic scattering of the Dirac  low energy ($\sim 1$ MeV) electron neutrinos ($\nu_e$'s)  on the polarized electron target \\(PET) in testing the time reversal symmetry violation (TRSV).   We consider a scenario in which the incoming $\nu_e$ beam is the  superposition of left chiral  (LC) and  right chiral (RC) states. LC $\nu_e$'s  interact mainly by the standard $V - A$ and small admixture of non-standard  scalar $S_L$, pseudoscalar $P_L$, tensor $T_L$ interactions, while RC ones are only detected  by the exotic $V + A$ and $S_R, P_R, T_R$ interactions. In addition, one assumes that the  spin  polarization vector of the initial $\nu_e$'s is turned aside  from  its momentum, and due to this the non-vanishing transversal component of the $\nu_e$ spin polarization appears.  
We compute the differential cross section as a function of the  recoil electron azimuthal angle and scattered electron energy, and show how the interference terms between standard $V-A$ and  exotic $S_R, P_R, T_R$ couplings depend on the various angular correlations among the transversal 
$\nu_e$ spin polarization, the polarization of 
the electron target, the incoming neutrino momentum and the outgoing electron momentum  in the limit of relativistic $\nu_e$. We illustrate   how   the maximal  value of recoil electrons azimuthal asymmetry and  the  asymmetry axis location of outgoing electrons depend on the azimuthal angle of the transversal component of the $\nu_e$ spin polarization, both for the time reversal symmetry conservation (TRSC) and TRSV.  
 Next, we display that the electron energy spectrum and polar angle distribution of the recoil electrons are also sensitive to the  interference terms between $V - A$ and  $S_R, P_R, T_R$ couplings, proportional to the T-even and T-odd angular correlations among the transversal 
$\nu_e$ polarization,  the electron polarization of the target, and the incoming $\nu_e$ momentum, respectively. 
 Our model-independent analysis is  carried out for the flavor  $\nu_e$'s.  To make such  tests 
feasible,  the intense (polarized) artificial $\nu_e$ source, PET  and  the appropriate detector measuring the directionality of the outgoing electrons, and/or the recoil electrons energy with a high resolution have to  be identified. 
\end{abstract}
\section{Introduction}
\label{sec1}
 One of the fundamental problems in  the neutrino physics is whether TRSV takes place  in purely leptonic processes at low energies (e. g. the  neutrino-electron elastic scattering (NEES). According to the standard electro-weak model (SM) 
\cite{SM,SM1,SM2,SM3,SM4}, the V and A couplings of LC $\nu_e$'s may only  participate in   NEES and the hermiticity conditions of interaction lagrangian require the real coupling constants. This means that there is no possibility of appearing  TRSV correlations in the differential cross section for the NEES, even when the electron target is polarized. The qualitative change emerges when the exotic scalar (S), tensor (T), pseudoscalar (P) and  (V+A) couplings of the interacting RC $\nu$'s beyond  the SM in addition to the standard V-A ones are introduced. 
 The presence of the exotic complex couplings together with PET may generate the non-vanishing T-even and T-odd angular correlations in the differential cross section. 
It is worth pointing out that the TRSV (equivalent CP violation in the case of CPT-invariant theory) is  observed in the decays of neutral
kaons and B-mesons \cite{CP,CP1,CP2}, and  described by a single phase of the Cabibbo-Kobayashi-Maskawa quark-mixing matrix (CKM) \cite{Kobayashi}. However, this  CP-violating phase  does not allow for explanation of the existing matter-antimatter asymmetry of universe  and new T-violating
 phases are needed \cite{barion}.  
It is important to note that the available experimental results  still do not rule out the scenarios with the exotic  S, T, P and V+A weak interactions  of RC $\nu$'s. The various non-standard gauge models including  exotic TRSV interactions, RC $\nu$'s, mechanisms explaining the origin of  parity violation and of fermion generations, masses, mixing and smallness of $\nu$ mass have been proposed. 
We mean, e. g., the left-right symmetric models (LRSM) \cite{Pati,Pati1,Pati2,Pati3,Pati4}, composite models \cite{Jodidio,CM,CM1}, models with extra dimensions (MED) \cite{Extra} and the unparticle  models (UP) \cite{unparticle,unparticle1,unparticle2,unparticle3,unparticle4,unparticle5,unparticle6,unparticle7,unparticle8,unparticle9,unparticle10,unparticle11,unparticle12}. Concerning the UP theory, it is noteworthy that in this scheme  $\nu$'s with  the different chiralities   can interact  with the spin-0 scalar, spin-1 vector, spin-2 tensor unparticle sectors and consequently one gets the amplitudes for  the 
low energy leptonic processes  in the form of  unparticle four-fermion contact interaction with  the non-standard S, T, P, V+A lorentz interactions.\\ In spite of experimental limitations and  lack of unambiguous indication of the non-standard model, there is a constant  necessity of increase of the  precision of  present tests at low energies, and on the other hand, it seems sensible to search for new tools sensitive to the linear effects from the  exotic complex  couplings of RC $\nu$'s, because the measurements of  these observables may shed some new light on the TRSV in the leptonic interactions. As it is known 
the future superbeam and neutrino factory projects  aim  at 
the tests of the CP violation in the lepton
sector, where simultaneously $\nu$ and $\overline{\nu}$ oscillations would be
measured \cite{Geer, Geer1}. Also other proposals of observables for the tests on  the TRSV in the
leptonic and semileptonic processes: the precise measurements of T-odd triple-correlations for the massive charged leptons \cite{neutron,Herczeg,Huber,Bodek,Mumm}, of    electric dipole moments of the neutron and atoms are worth noticing \cite{NDM,NDM1,NDM2,NDM3,NDM4}. Till now,  all the evidence  is consistent with the TRSC scenario. 
\\
Our considerations  show that  NEES on  PET 
offers new scientific opportunities for the studies on the TRSV in the leptonic reactions. From the perspective of the main goals of this paper, it is essential to  mention the recent tests confirming  the possibility of realizing the polarized target crystal of $Gd_{2}SiO_{5}$ doped with Cesium \cite{INFN}, as suggested in \cite{Misiaszek}. The concepts of using   PET to probe the neutrino magnetic moments,  the flavor composition of  (anti)neutrino beam, axions, spin-spin interaction in gravitation \cite{PET,PET1,PET2,PET3,PET4,PET5,PET6,PET7} are also  worth noting. \\
In this study, we focus  on the elastic  scattering of  low energy  Dirac $\nu_e$'s  on  PET. We   show in a  model-independent way how the admixture of the exotic S, T, P, V+A complex couplings of RC $\nu_e$'s  in addition to the standard V, A  real couplings  of LC ones affects on the azimuthal distribution and asymmetry of the recoil electrons, polar distribution of scattered electrons and their energy spectrum, and consequently  on  the possibility of  TRSV in the relativistic    $\nu_e$ limit. 
Our studies are made for the flavor-eigenstate (current) Dirac $\nu_e$'s and  when  the monochromatic $\nu_e$ source is deployed at a  near distance from the detector.  
We analyze the various scenarios assuming that the hypothetical detector is able to measure  both the azimuthal angle $\phi_{e}$ and polar angle $\theta_e$ of the recoil electrons, and/or also the energy of the outgoing electrons with a high resolution.  In order to compute the expected effects, we  use the experimental values of standard couplings: $c_{V}^{L}= 1 + (-0.04 \pm 0.015), c_{A}^{L}= 1+ \\(-0.507 \pm 0.014)$   \cite{Data}. 

\section{Elastic scattering of Dirac electron neutrinos  on polarized electrons - basic assumptions} 
\label{sec2}
We  assume that the incoming   Dirac $\nu_{e}$  beam is generated by  the monochromatic low energy ($\sim 1 MeV$) and  polarized source ( $\nu_{e}$ emitter with  a high intensity). Let us remind that the $^{51}Cr$ unpolarized emitter with  a  high activity  $\sim 370 \; PBq$  in the SOX experiment  \cite{SoxBell} (Short distance Oscillation with bor\-eX\-ino) at the Borexino detector is planed to search for among other the sterile $\nu_e$'s 
\cite{sterile,sterile1,sterile2,sterile3,sterile4,sterile5,sterile6}. Moreover, one supposes that the initial $\nu_{e}$   flux is  the superposition of LC states detected mainly by the standard $V - A$ and small admixture of non-standard  scalar $S_L$, pseudoscalar $P_L$, tensor $T_L$ interactions, while RC ones interact only  by the exotic $V + A$ and $S_R, P_R, T_R$ interactions. Additionally, one admits that the initial $\nu_{e}$ beam  has   the  spin  polarization vector turned aside from  its  momentum, and in this way  the non-vanishing transversal components of the spin polarization appear. 
In order to illustrate the possibility of producing the $\nu_{e}$ beam with the non-zero transversal spin polarization, we refer to the ref. \cite{CMS2003}, where the muon capture by proton as the production process of L-R chiral superposition has been considered. In the next studies, the other sources are going to be analyzed. It should be stressed  that when the admixture of RC $\nu$'s in addition to the LC ones  in the polarized $\nu$ source is admitted and the production plane is assigned, the $\nu$ spin polarization vector may acquire the transversal components, potentially giving both T-even and T-odd effects. These transversal $\nu$ polarizations  consist only of the interferences between the $(V,A)_{L}$ and $(S, T, P)_{R}$ couplings and do not vanish in the relativistic $\nu$ limit.  We have completely different situation for the longitudinal  $\nu$  polarization, where  all the  interferences between the $V-A$ and $(S, T, P)_{R}$ interactions are strongly suppressed by $\nu$ mass. It means that only the squares of exotic RC  couplings (at most  the interferences within exotic couplings) and of standard LC ones  may generate the possible effect. 
 The amplitude for the   $\nu_{e}  e^{-}$ scattering  takes the form:
\beq \label{ampD} M^{D}_{\nu_{e} e^{-}}
&=&
\frac{G_{F}}{\sqrt{2}}\{(\overline{u}_{e'}\gamma^{\alpha}(c_{V}^{L}
- c_{A}^{L}\gamma_{5})u_{e}) (\overline{u}_{\nu_{e'}}
\gamma_{\alpha}(1 - \gamma_{5})u_{\nu_{e}})\nonumber\\ 
&& \mbox{} + (\overline{u}_{e'}\gamma^{\alpha}(c_{V}^{R}
+ c_{A}^{R}\gamma_{5})u_{e}) (\overline{u}_{\nu_{e'}}
\gamma_{\alpha}(1 + \gamma_{5})u_{\nu_{e}})  \\
&  & \mbox{} +
c_{S}^{R}(\overline{u}_{e'}u_{e})(\overline{u}_{\nu_{e'}}
(1 + \gamma_{5})u_{\nu_{e}}) \nonumber\\
&  & \mbox{} +
c_{P}^{R}(\overline{u}_{e'}\gamma_{5}u_{e})(\overline{u}_{\nu_{e'}}
\gamma_{5}(1 + \gamma_{5})u_{\nu_{e}}) \nonumber\\
&& \mbox{} +
\frac{1}{2}c_{T}^{R}(\overline{u}_{e'}\sigma^{\alpha \beta}u_{e})(\overline{u}_{\nu_{e'}}
\sigma_{\alpha \beta}(1 + \gamma_{5})u_{\nu_{e}})\nonumber\\
&&\mbox{} +
c_{S}^{L}(\overline{u}_{e'}u_{e})(\overline{u}_{\nu_{e'}}
(1 - \gamma_{5})u_{\nu_{e}}) \nonumber\\
&&\mbox{} + c_{P}^{L}(\overline{u}_{e'}\gamma_{5}u_{e})(\overline{u}_{\nu_{e'}}
\gamma_{5}(1 - \gamma_{5})u_{\nu_{e}}) \nonumber\\
&& \mbox{} +
\frac{1}{2}c_{T}^{L}(\overline{u}_{e'}\sigma^{\alpha \beta}u_{e})(\overline{u}_{\nu_{e'}}
\sigma_{\alpha \beta}(1 - \gamma_{5})u_{\nu_{e}})
\}\nonumber,  
 \eeq
where $G_{F} = 1.1663788(7)\times
10^{-5}\,\mbox{GeV}^{-2} (0.6 \; ppm)$ \cite{Mulan} is the Fermi constant. 
The coupling constants are
denoted with the superscripts $L $ and $R $ as $c_{V}^{L, R} $, 
$c_{A}^{L, R}$, $c_{S}^{R, L}$, $c_{P}^{R, L}$,  $c_{T}^{R, L}$ respectively to the incoming $\nu_{e}$ 
of left- and right-handed chirality.  Because we admit  the TRSV,  the non-standard coupling constants $c_{S}^{R, L}$, $c_{P}^{R, L}$,  $c_{T}^{R, L}$  are the complex numbers denoted as 
$c_S^R = |c_S^R|e^{i\,\theta_{S,R}}$, $c_S^L = |c_S^L|e^{i\,\theta_{S,L}}$, etc.  
Moreover, the relations between the exotic couplings,  
$c_{S, T, P}^{ L}=c_{S, T, P}^{*R}$ appearing at the level of interaction lagrangian should be taken into account. 
It manifests  the lack of  dependence of the square terms coming from the $S, T, P$ interactions in the cross section  on the longitudinal $\nu_{e}$  polarization $\mbox{\boldmath $\hat{\eta}_{\nu}$}\cdot\hat{\bf q}$.
The general formula for the differential cross section with the  dependence on the azimuthal angle of  outgoing electron momentum, when 
$\mbox{\boldmath $\hat{\eta}_{e}$} \perp {\bf \hat{ q}}$, is presented in the appendix. Calculations are carried out with  the use of the covariant 
projectors   for the incoming $\nu_{e}$'s (including both the longitudinal and transverse components of the spin polarization) in the relativistic limit   and  for the polarized target-electrons, respectively \cite{Michel}. 
\begin{figure}
\begin{center}
\includegraphics[scale=.65]{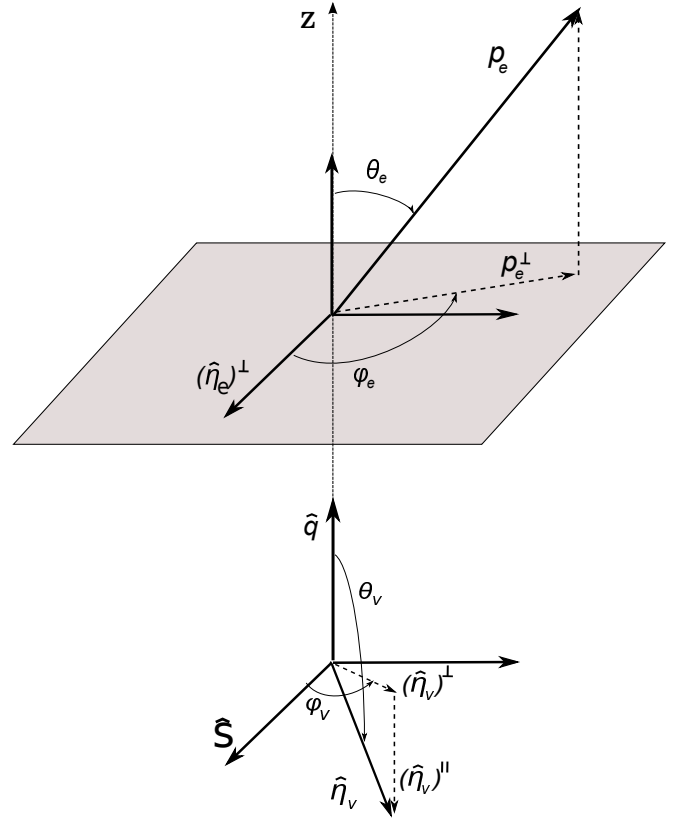}
\end{center}
\caption{Production plane of the 
$\nu_{e}$ beam is   spanned by the polarization unit vector $\mbox{\boldmath $\hat{S}$}$
 of  source  and  the $\nu_{e}$  LAB momentum unit vector ${\bf \hat{q}}$. Reaction  plane is spanned by  $\hat{\bf q}$ and  the transverse electron
polarization vector of target $\mbox{(\boldmath $ \hat{\eta}_{e})^\perp $}$ (due to $\mbox{\boldmath $\hat{\eta}_{e}$} \perp {\bf \hat{ q}}$) for $\nu_{e} + e^{-}\rightarrow \nu_{e} +e^{-}$.  
$\theta_{e}$ is the polar angle between ${\bf \hat{q}}$ and 
 the unit vector $ \hat{\bf p}_{e}$ of recoil electron momentum. $\phi_{e}$ is the 
angle between $\mbox{(\boldmath $ \hat{\eta}_{e})^\perp $}$ and the transversal component of  outgoing electron momentum $ (\hat{\bf p}_{e})^{\perp}$.$\mbox{\boldmath $\hat{\eta}_{\nu}$}=(\sin\theta_\nu \,\cos\phi_{\nu}, \sin\theta_\nu \,\sin\phi_{\nu}, \cos\theta_\nu)$.    
  \label{Fig1D}}
\end{figure} 

\section{Azimuthal distribution and asymmetry of recoil electrons} 

In this section, we analyze the possibility of using the  azimuthal distribution of recoil electrons for the investigation of TRSV in  the $\nu$ elastic scattering on  PET.  According to the SM, the mentioned azimuthal distribution  has a local maximum at $\Phi=\pi/2$ as it is illustrated in the Fig.2 and Fig.3, respectively. 
The Fig. 2 is the polar plot of $d^2 \sigma/d\phi_e d\theta_e=\left(d^2 \sigma/d\phi_e d \,y\right) \cdot \left(d \,y/d\theta_e\right)$ as a function of $\phi_e$ for the assigned values of $\theta_e$.  The Fig. 3 is the polar plot of $d^2 \sigma/d\phi_e d \,y$ as a function of $\phi_e$ for the assigned values of $y$.  
These two plots reveal the up-down azimuthal  symmetry measured with respect to $\Phi=0$ and the left-right azimuthal asymmetry with the asymmetry axis  directed along  $\Phi=\pi/2$, clearly visible  for $d^2 \sigma/d\phi_e d\theta_e$. Moreover, it is important to stress that in the case of the standard $V-A$ interaction, the asymmetry axis is fixed at $\Phi=\pi/2$ and is independent of the variations of $y$, $\theta_e$, $E_{\nu}$ and the standard $c_V^L$, $c_A^L$ couplings values but the degree of the asymmetry can change. 
Usually to quantify the azimuthal asymmetry one makes use of the asymmetry function (see Appendix 2 for the definitions).  Fig.4 displays the maximal values of azimuthal asymmetries $A_y(\Phi=\pi/2)$ and $A_{\theta_e}(\Phi=\pi/2)$  as functions of $y$ and $\theta_e$ for the standard $V-A$ interaction. In  both cases the maximal values of the $A_y$ and $A_{\theta_e}$ are equal to $0.0794$, and achieved at $y^{max} \approx 0.5$ and $\theta_{e}^{max} \approx \pi/6$, respectively. 
\label{sec3}
\begin{figure}
\begin{center}
\includegraphics[scale=.6]{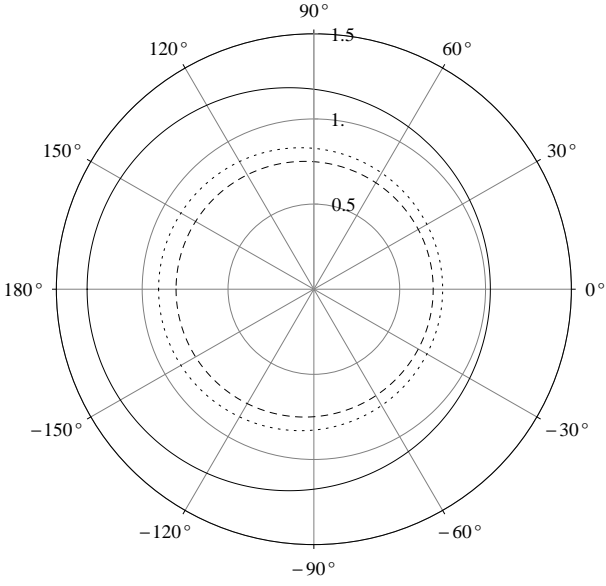} 
\end{center}
\caption{ Dependence of $d^2 \sigma/d\phi_e d\,\theta_e$ on $\phi_e$ for  the standard V-A interaction, $E_\nu=1\,MeV$:  $\theta_e=\pi/12$ (dotted line), $\theta_e=\pi/6$ (solid line), $\theta_e=\pi/3$ (dashed line). \label{Fig2}}
\end{figure}
\begin{figure}
\begin{center}
\includegraphics[scale=.6]{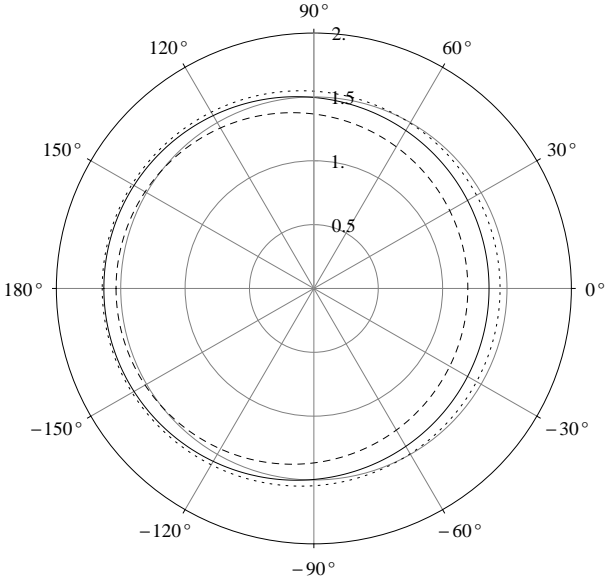}
\end{center}
\caption{Dependence of $d^2 \sigma/d\phi_e d\,y$ on $\phi_e$ for  the standard V-A interaction, $E_\nu=1\,MeV$:  $y=0.1$ (dotted line), $y=0.2$ (solid line), $y=0.5$ (dashed line).  \label{Fig3}}
\end{figure}
\begin{figure}
\begin{center}
\includegraphics[scale=.4]{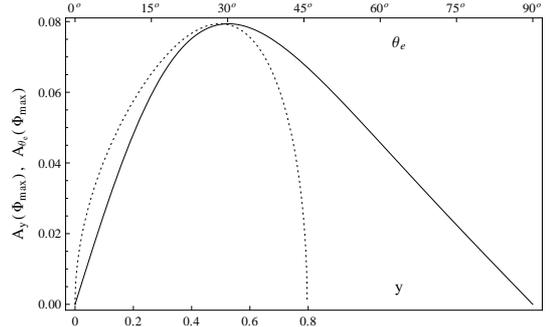}
\end{center}
\caption{Standard V-A interaction, $E_\nu=1\,MeV$. Plot of the azimuthal asymmetry functions: 
$A_y(\Phi=\pi/2)$ as a function y (solid line) and $A_{\theta_e}(\Phi=\pi/2)$ as a function of $\theta_e$ (dotted line).  \label{Fig4}}
\end{figure}
\begin{figure}
\begin{center}
\includegraphics[scale=.34]{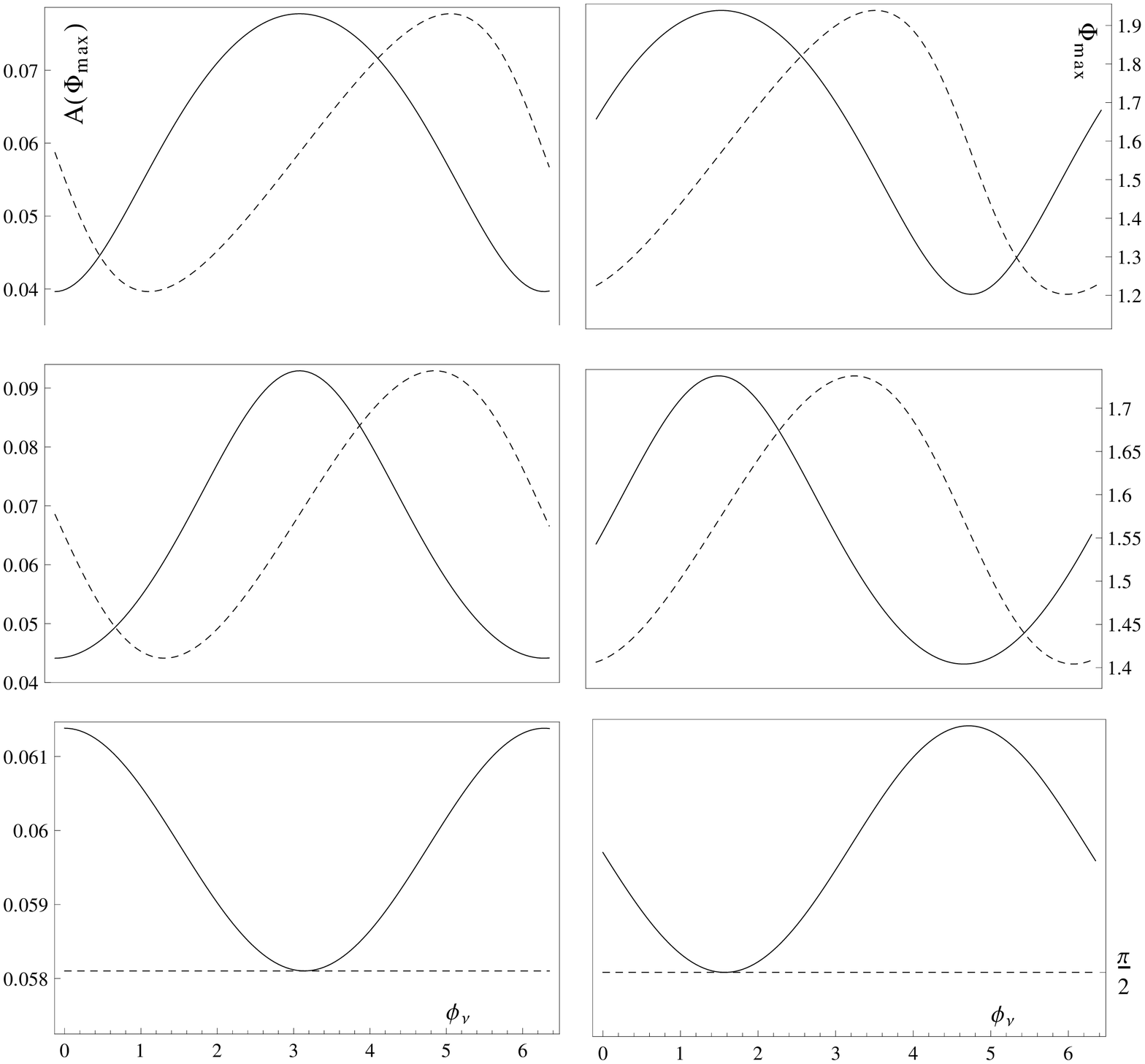}
\end{center}
\caption{ Dependence of $A(\Phi_{max})$ on $\phi_\nu$  (solid line) and $\Phi_{max}$ on $\phi_\nu$ (dashed line), for $\mbox{\boldmath $\hat{\eta}_{\nu}$}\cdot\hat{\bf q}=-0.95, E_\nu=1\,MeV$. TRSC: Upper left plot for the case of   $V-A $ and $T_R$ when $|c_T^R|=0.2, \theta_{T,R}=0$; Middle left plot for the combination of $V-A $ with $S_R$ when $|c_S^R|=0.2, \theta_{S,R}=0$; Lower left plot for the case of $V-A $ with $P_R$ when $|c_P^R|=0.2, \theta_{P,R}=0$. TRSV:  Upper right plot for the scenario with  $V-A $ and $T_R$ when $|c_T^R|=0.2, \theta_{T,R}=\pi/2$;  Middle right plot for the case of $V-A $ and $S_R$ when $|c_S^R|=0.2, \theta_{S,R}=\pi/2$; Lower right plot for the combination of $V-A $ with $P_R$ when $|c_P^R|=0.2, \theta_{P,R}=\pi/2$. \label{Fig5}}
\end{figure}
\begin{figure}
\begin{center}
\includegraphics[scale=.34]{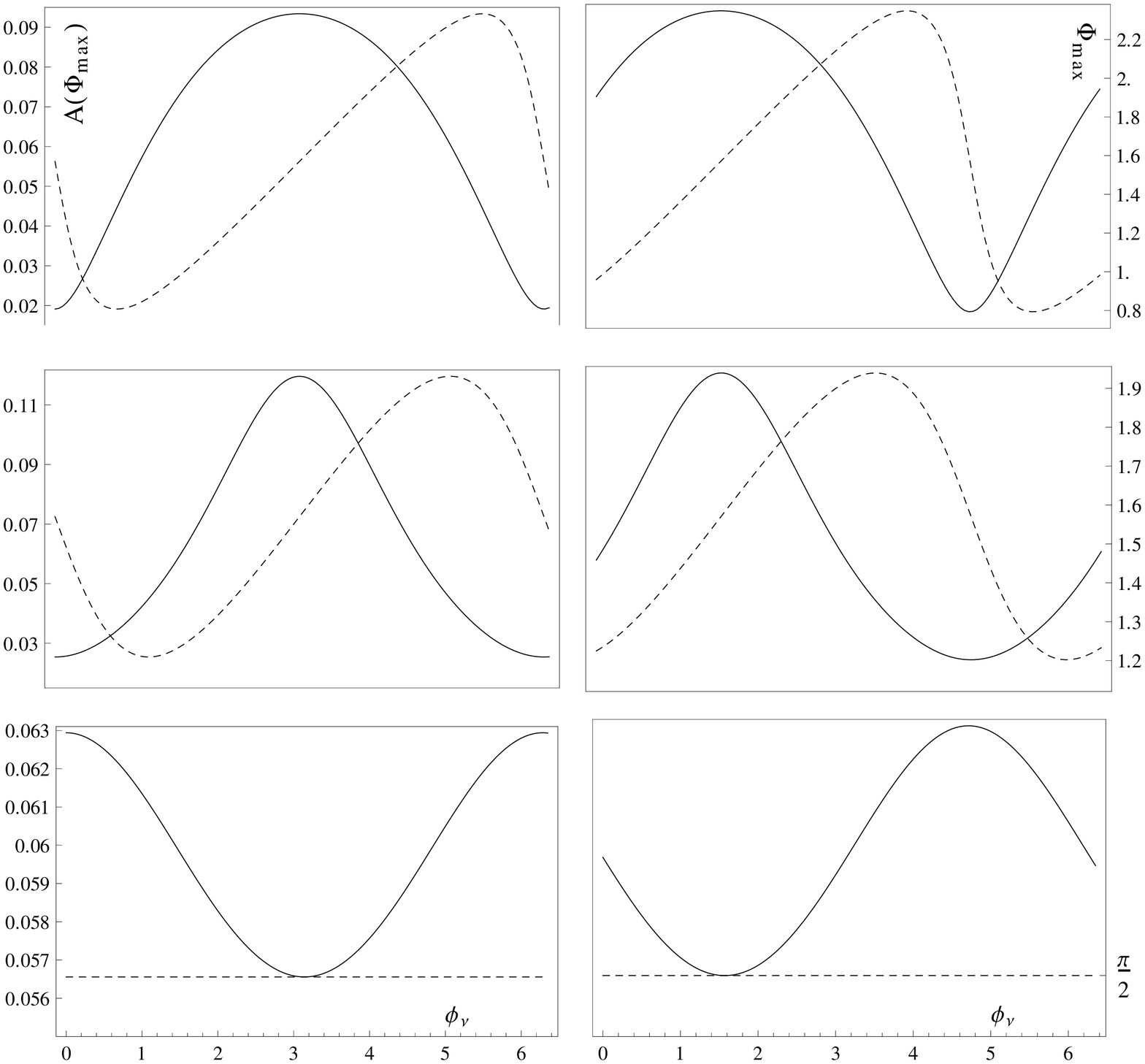}
\end{center}
\caption{Dependence of $A(\Phi_{max})$ on $\phi_\nu$  (solid line) and $\Phi_{max}$ on $\phi_\nu$ (dashed line), for $\theta_\nu=\pi/2$, $E_\nu=1\,MeV$. TRSC: Upper left plot for the case of   $V-A $ and $T_R$ when $|c_T^R|=0.2, \theta_{T,R}=0$; Middle left plot for the combination of $V-A $ with $S_R$ when $|c_S^R|=0.2, \theta_{S,R}=0$; Lower left plot for the case of $V-A $ with $P_R$ when $|c_P^R|=0.2, \theta_{P,R}=0$. TRSV: Upper right plot for the scenario with  $V-A $ and $T_R$ when $|c_T^R|=0.2, \theta_{T,R}=\pi/2$;  Middle right plot for the case of $V-A $ and $S_R$ when $|c_S^R|=0.2, \theta_{S,R}=\pi/2$; Lower right plot for the combination $V-A $ with $P_R$ when $|c_P^R|=0.2, \theta_{P,R}=\pi/2$.  \label{Fig6}}
\end{figure}
In order to illustrate how the phase of given exotic coupling and the azimuthal angle of $\mbox{\boldmath $(\hat{\eta}_{\nu})^{ \perp}$}$ affect the  azimuthal asymmetry of recoil electrons, and consequently to
hint to the possibility of TRSV, we  present the explicit form of the $A(\Phi)$ asymmetry function for the scenario with $V-A$ and $S_R$ interactions:
\beq 
\label{interVASR}
A_{V-A}^{S,R}(\Phi) & = & 3 \sqrt{m_e (2 E_{\nu} + m_e)} \bigg[-(c_{A}^{L} - c_V^L) (3 c_A^L E_{\nu} + c_V^L E_{\nu}  \nonumber\\ 
&& + 2 c_A^L m_e) \sin^2{\frac{\theta_{\nu}}{2}} \sin{\Phi}  \\ 
&& + |c_S^R| c_V^L (2 E_{\nu} + m_e) \sin{(\theta_{S,R} - \Phi + \phi_{\nu})}\bigg]/ \nonumber \\
&&
\bigg[4 (|c_S^R|^2 + 4 (c_A^L{}^2 + c_A^L c_V^L + c_V^L{}^2)) E_{\nu}^2 \nonumber \\ 
&& + 12 (2 c_A^L{}^2 + |c_S^R|^2 + c_A^L c_V^L + c_V^L{}^2) E_{\nu} m_e \nonumber \\ 
&& + 3 (3 c_A^L{}^2 + 2 |c_S^R|^2 + c_V^L{}^2) m_e^2 \nonumber \\
&& - (4 c_A^L c_V^L E_{\nu} (4 E_{\nu} + 3 m_e) + c_A^L{}^2 (4 E_{\nu} + 3 m_e)^2 \nonumber \\
&& +     c_V^L{}^2 (16 E_{\nu}^2 + 12 E_{\nu} m_e + 3 m_e^2)) \cos{\theta_{\nu}} \nonumber \\
&& - |c_S^R| (6 (c_A^L + c_V^L) E_{\nu}^2 + (13 c_A^L + 5 c_V^L) E_{\nu} m_e \nonumber \\
&& +  6 c_A^L m_e^2) \cos{(\theta_{S,R} + \phi_{\nu})}\bigg]\nonumber. 
\eeq 
We see that if $|c_S^R|\,\sin{(\theta_{S,R} + \phi_{\nu})} = 0$ then the local extremum of $A_{V-A}^{S,R}(\Phi)$ is at $\Phi=\pi/2$. Assuming no $T_R$ and $P_R$ interactions it follows that any departure from the $\Phi_{max}=\pi/2$ orientation of the asymmetry axis signalizes the presence of the exotic $c_S^R$ interaction. Moreover, if the location of the asymmetry axis is sensitive to the relative  orientation of PET and $\mbox{\boldmath $(\hat{\eta}_{\nu})^{ \perp}$}$ one can conclude that 
$\theta_{S,R}+\phi_{\nu}\neq 0$; ideally, experimental control of $\phi_{\nu}$ would give an opportunity 
to measure $\theta_{S,R}$. For example, $\theta_{S,R}$ can be determined as $-\phi_{\nu}$ for that value  of $\phi_{\nu}$ which fixes the asymmetry axis at the standard $\pi/2$ location. 
The similar regularity holds for the case of $V-A$ and $T_R$ interactions, i.e. when  $|c_T^R|\,\sin{(\theta_{T,R}+ \phi_{\nu})} = 0$ then the local extremum of $A_{V-A}^{T,R}(\Phi)$ must be at $\Phi=\pi/2$. 
For the variant with $V-A$ and $P_R$ couplings the situation is different: in this case  $\Phi_{max}=\pi/2$ independently of the coupling $c_P^R$ and the azimuthal angle $\phi_{\nu}$. 
\par The diagrams on Figs. 5, 6 show dependence of $A(\Phi_{max})$ (solid lines) and $\Phi_{max}$ (dashed lines) on $\phi_\nu$ for the various scenarios
with TRSC (left plots) and TRSV (right plots). 

\section{Spectrum of recoil electrons and polar angle distribution of scattered electrons} 
\label{sec4}

In this section, we indicate the usefulness of both polar angle distribution and spectrum of recoil electrons in testing the TRSV phenomenon. Let us stress that the probed scenarios correspond to the laboratory differential cross section integrated over   $\phi_{e}$. We see that this independence of $\phi_{e}$ does not eliminate all the  interference terms between the standard and exotic $(S,T,P)_R$ couplings, and in this way there is still the possibility of detecting the TRSV by the precise measurement of these observables. Fig.7 shows the dependence of $d\sigma/d\theta_e=\left(d\sigma/d \,y\right) \, \left(d \,y/d\theta_e \right)$ on $\theta_e$ for the various scenarios. Upper plot concerns  TRSC ($\theta_{T,R}=\theta_{S,R}=\theta_{P,R}=0$) case, while lower one corresponds to  TRSV ($\theta_{T,R}=\theta_{S,R}=\theta_{P,R}=\pi/2$). Fig. 8  displays the same dependence as the Fig. 7, but for the pure contribution from the $\mbox{\boldmath $(\hat{\eta}_{\nu})^{ \perp}$}$, i. e.  when $\theta_\nu=\pi/2$. 
The significant departure from the standard prediction  in the polar angle distribution of recoil electrons  for the scenario with $V-A$ and $T_R$ interactions  can be noticed. For two remaining cases the differences are much smaller.  The dashed lines in Fig. 7 correspond to the case of TRSC (upper plot) with  $\theta_{e}^{max}(T_R)=34.3^{\circ}$ and TRSV (lower plot) with  $\theta_{e}^{max}(T_R)= 33.3^{\circ}$, respectively. The similar regularity for the extreme situation with $\theta_\nu=\pi/2$ is seen in Fig. 8. Figs. 9, 10 are the plots of $d\sigma/d \,y$ as a function of $y$ depicted with the similar assumptions as for the Figs. 7, 8. \\
We  present the  recoil electrons energy spectrum  $d\sigma/d \,y$ for the scenario with $V-A$ and $S_R$ interactions  to illustrate the impact of   phase of  exotic coupling and azimuthal angle of $\mbox{\boldmath $(\hat{\eta}_{\nu})^{ \perp}$}$ on the possibility of TRSV:
\beq
\label{fgS}
\bigg(\frac{d\sigma}{d y}\bigg)_{V-A,S} & = & \bigg(\frac{d\sigma}{d y}\bigg)_{V-A} \\ \nonumber
&&
\hspace{-1cm}+\; B\,\bigg[ |c_S^R|^2\, f_{S}(y) + |c_S^R|\cos{(\theta_{S,R}+\phi_{\nu})} g_{S}(y)\bigg],
\eeq
with the $y$-dependent coefficients
\beq
f_{S}(y) & = & 2\,y\,(\frac{2 m_e}{E_{\nu}}+y) \\ \nonumber
g_{S}(y) & = & y \left[(-2(c_A^L + c_V^L) + y (c_A^L-c_V^L)\frac{m_e}{E_{\nu}} - 
4 c_A^L \frac{m_e}{E_{\nu}}\right].
\eeq
The similar decomposition but with different coefficients $f$, $g$ holds for the $V-A$ and $T_R$, also $V-A$ and $P_R$ interactions. Let us remark that the  low energy region of  recoil electrons spectrum largely deviates from the standard expectation for the $V-A$ and $T_R$ couplings (dashed line in  Figs. 9, 10). The cases of $V-A $ with $S_R$ and $V-A $ with $P_R$ indicate the relatively small deviation for the higher outgoing electrons energy.  
\begin{figure}
\begin{center}
\includegraphics[scale=.5]{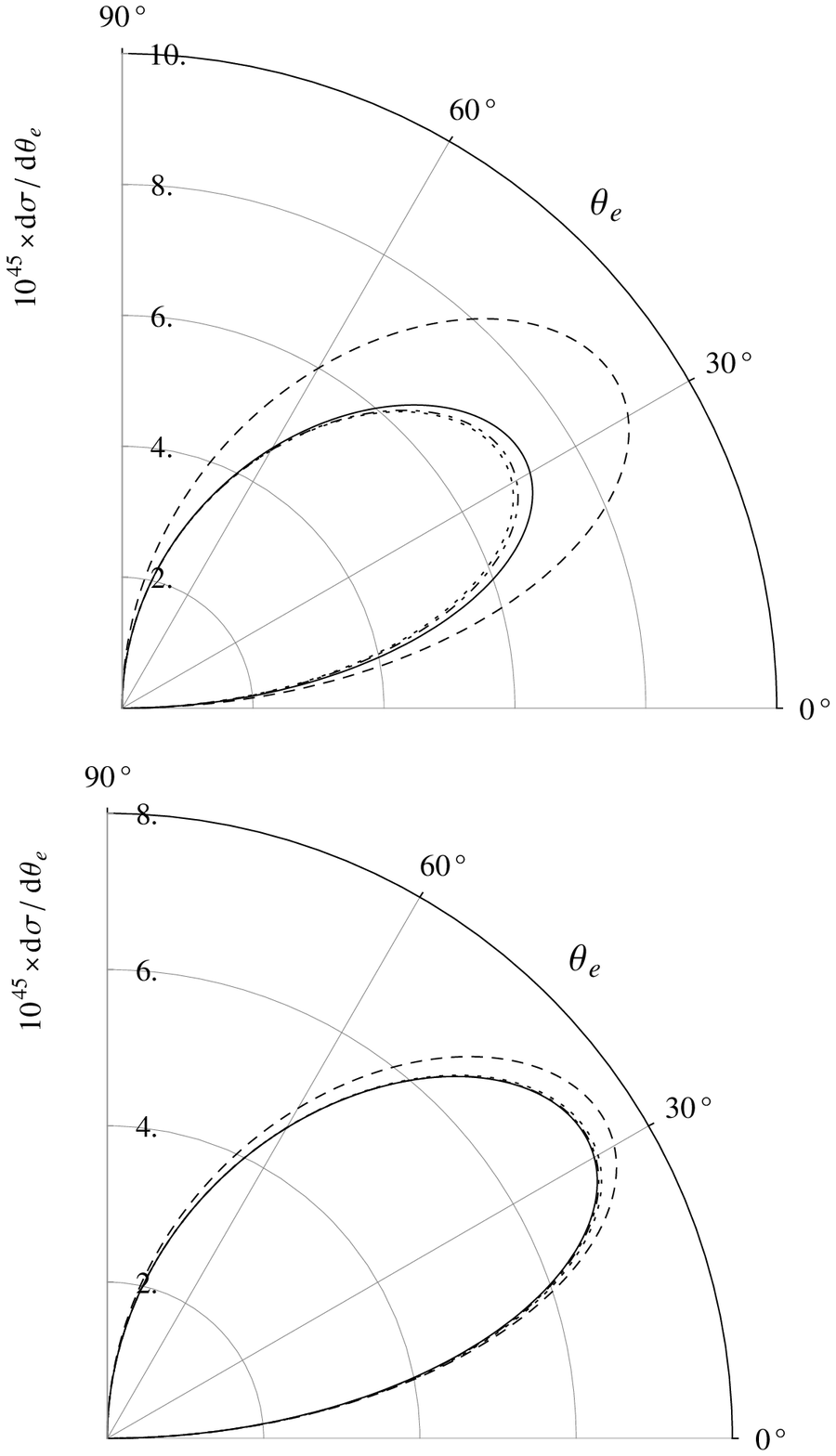}
\end{center}
\caption{Dirac $\nu_e$. Plot of $d\sigma/d\theta_e$ as a function of $\theta_e$ for $\mbox{\boldmath $\hat{\eta}_{\nu}$}\cdot\hat{\bf q}= - 0.95$, $E_\nu=1\,MeV$.
Upper plot for TRSC: standard $V-A$ interaction (solid line);  the combination of $V-A $ and $T_R$ when $|c_T^R|=0.2, \theta_{T,R}=0$ (dashed line); the case of $V-A $ and $S_R$ when $|c_S^R|=0.2, \theta_{S,R}=0$ (dotted line); $V-A $ with $P_R$ when $|c_P^R|=0.2, \theta_{P,R}=0$ (dashed-dotted line).
Lower plot for TRSV: standard $V-A$ interaction (solid line);  the combination of $V-A $ and $T_R$ when $|c_T^R|=0.2, \theta_{T,R}=\pi/2$ (dashed line); the case of $V-A $ and $S_R$ when $|c_S^R|=0.2, \theta_{S,R}=\pi/2$ (dotted line); $V-A $ with $P_R$ when $|c_P^R|=0.2, \theta_{P,R}=\pi/2$ (dashed-dotted line).  \label{Fig7}}  
\end{figure}
\begin{figure}
\begin{center}
\includegraphics[scale=.48]{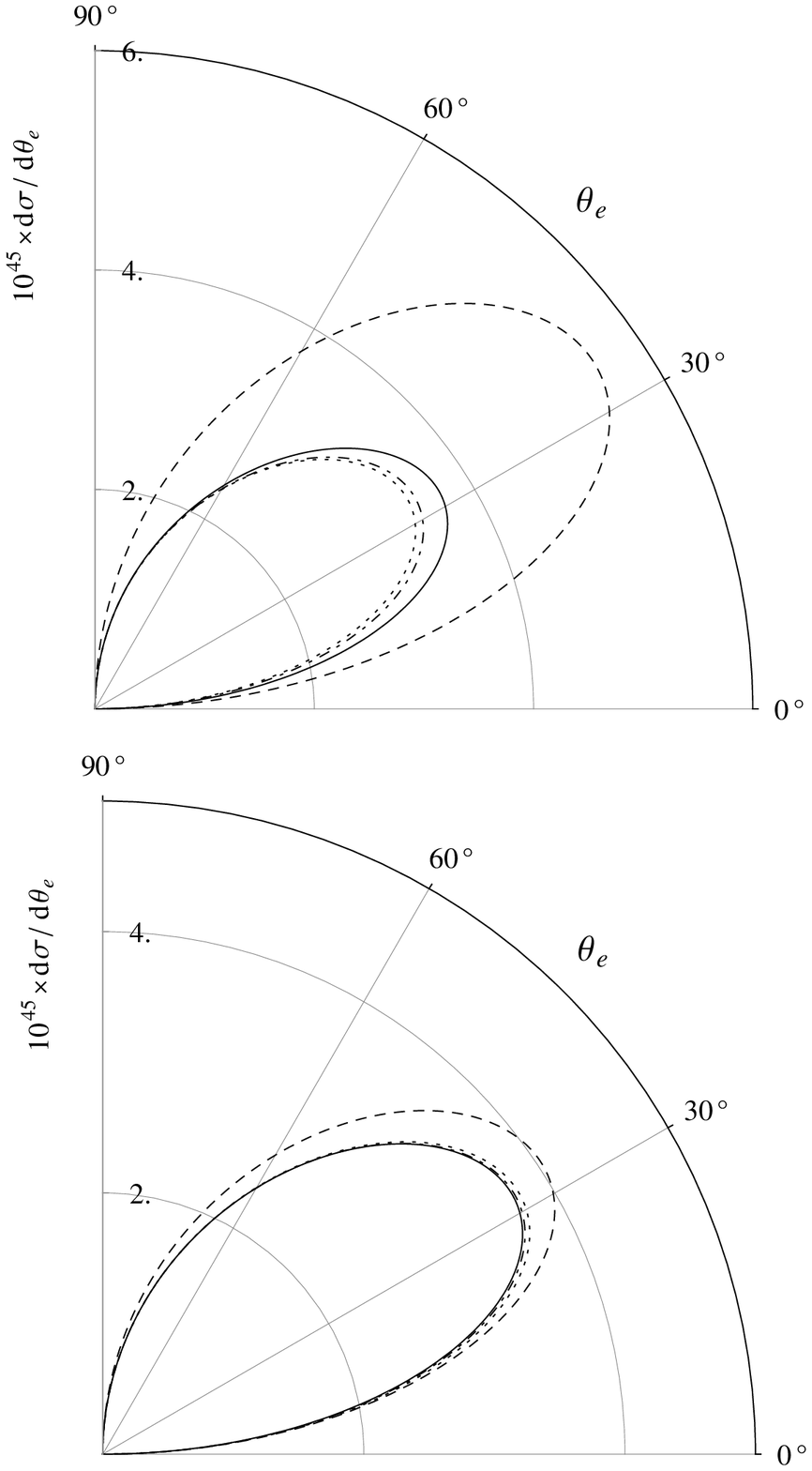}
\end{center}
\caption{Dirac $\nu_e$. Plot of $d\sigma/d\theta_e$ as a function of $\theta_e$ for $\theta_\nu=\pi/2$, $E_\nu=1\,MeV$. 
Upper plot for TRSC: standard $V-A$ interaction (solid line);  the combination of $V-A $ and $T_R$ when $|c_T^R|=0.2, \theta_{T,R}=0$ (dashed line); the case of $V-A $ and $S_R$ when $|c_S^R|=0.2, \theta_{S,R}=0$ (dotted line); $V-A $ with $P_R$ when $|c_P^R|=0.2, \theta_{P,R}=0$ (dashed-dotted line).
Lower plot for TRSV: standard $V-A$ interaction (solid line);  the combination  of $V-A $ and $T_R$ when $|c_T^R|=0.2, \theta_{T,R}=\pi/2$ (dashed line); the case of $V-A $ and $S_R$ when $|c_S^R|=0.2, \theta_{S,R}=\pi/2$ (dotted line); $V-A $ with $P_R$ when $|c_P^R|=0.2, \theta_{P,R}=\pi/2$ (dashed-dotted line).  \label{Fig8}}
\end{figure}
\begin{figure}
\begin{center}
\includegraphics[scale=.55]{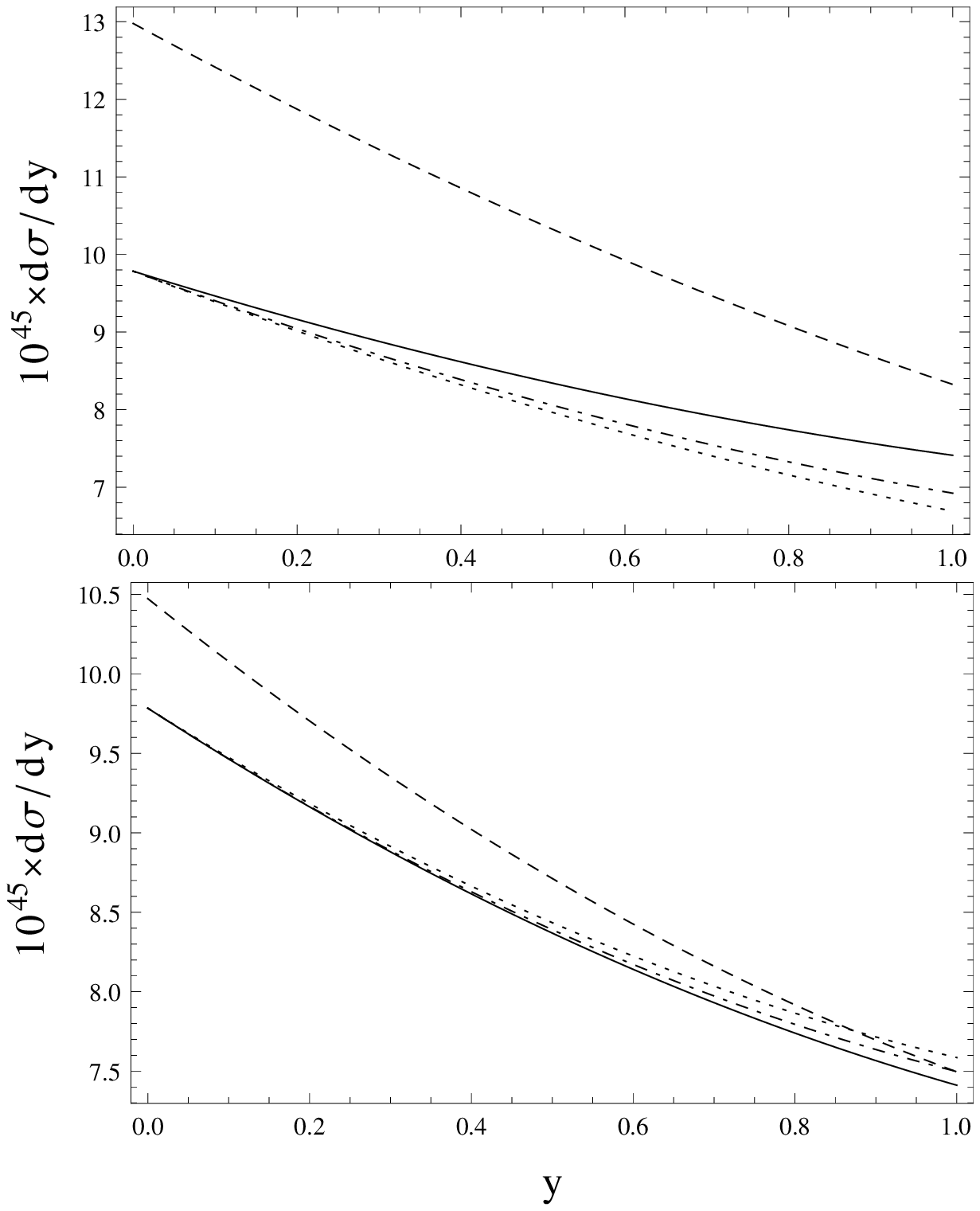}
\end{center}
\caption{ Dependence of  $d\sigma/d \,y$ on $y$  for $\mbox{\boldmath $\hat{\eta}_{\nu}$}\cdot\hat{\bf q}=-0.95, E_\nu=1\,MeV$. 
Upper plot for TRSC: standard $V-A$ interaction (solid line);  the combination of $V-A $ and $T_R$ when $|c_T^R|=0.2, \theta_{T,R}=0$ (dashed line); the case of $V-A $ and $S_R$ when $|c_S^R|=0.2, \theta_{S,R}=0$ (dotted line); $V-A $ with $P_R$ when $|c_P^R|=0.2, \theta_{P,R}=0$ (dashed-dotted line).
Lower plot for TRSV: standard $V-A$ interaction (solid line);  the combination of $V-A $ and $T_R$ when $|c_T^R|=0.2, \theta_{T,R}=\pi/2$ (dashed line); the case of $V-A $ and $S_R$ when $|c_S^R|=0.2, \theta_{S,R}=\pi/2$ (dotted line); $V-A $ with $P_R$ when $|c_P^R|=0.2, \theta_{P,R}=\pi/2$ (dashed-dotted line).   \label{Fig9}}
\end{figure}
\begin{figure}
\begin{center}
\includegraphics[scale=.55]{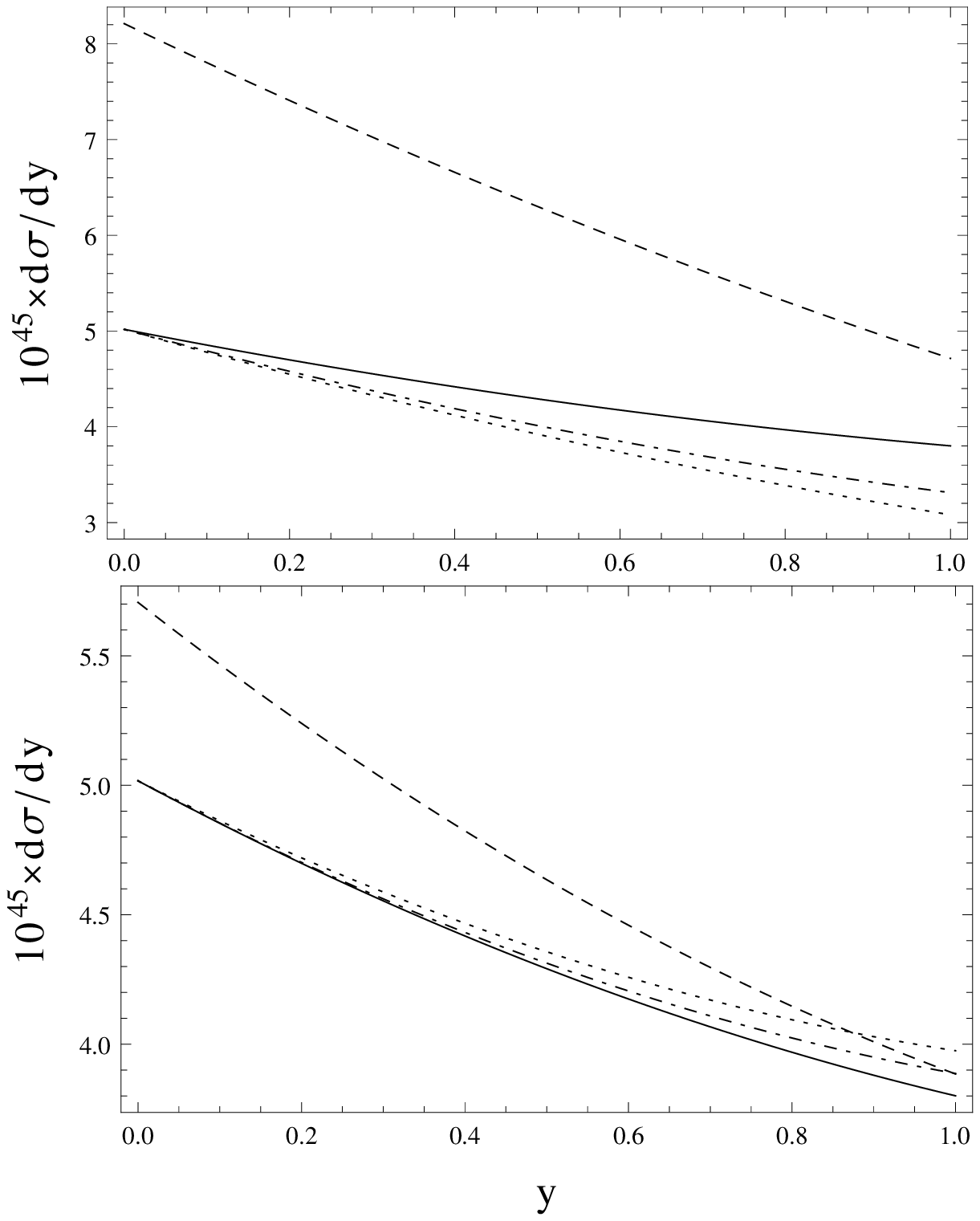}
\end{center}
\caption{Dependence of   $d\sigma/d \,y$ on $y$ for $\theta_\nu=\pi/2$, $E_\nu=1\,MeV$.  
Upper plot for TRSC: standard $V-A$ interaction (solid line); the combination of $V-A $ and $T_R$ when $|c_T^R|=0.2, \theta_{T,R}=0$ (dashed line); the case of $V-A $ and $S_R$ when $|c_S^R|=0.2, \theta_{S,R}=0$ (dotted line); $V-A $ with $P_R$ when $|c_P^R|=0.2, \theta_{P,R}=0$ (dashed-dotted line).
Lower plot for TRSV: standard $V-A$ interaction (solid line);  the combination of $V-A $ and $T_R$ when $|c_T^R|=0.2, \theta_{T,R}=\pi/2$ (dashed line); the case of $V-A $ and $S_R$ when $|c_S^R|=0.2, \theta_{S,R}=\pi/2$ (dotted line); $V-A $ with $P_R$ when $|c_P^R|=0.2, \theta_{P,R}=\pi/2$ (dashed-dotted line).  \label{Fig10}}
\end{figure}
\section{Conclusions}
\label{concl}

We have shown that  PET may be the useful tool for the detection  of the non-standard couplings of  RC interacting 
$\nu_e$'s and the effects of TRSV caused by the triple correlations present in the differential cross section for NEES. 
First, according to the SM prediction the left-right azimuthal asymmetry  of the 
recoil electrons  has the maximal value  at $\Phi=\pi/2$ and the location of asymmetry axis is fixed, Figs. 2, 3, 4. 
If  exotic $(S, T, P)_{R}$ complex couplings are admitted in NEES on  PET, both the  magnitude $A(\Phi_{max})$ and  axis $\Phi_{max}$ of the azimuthal asymmetry may change due to the non-vanishing interferences between the $V-A$ and $(S, T, P)_{R}$ proportional to the TRSC and TRSV correlations, Figs. 5, 6. 
 This departure from the standard prediction is mainly caused by the dependence of azimuthal asymmetry on the azimuthal angle $\phi_\nu$ connected with $\mbox{\boldmath $(\hat{\eta}_{\nu})^{ \perp}$}$ as it is shown in the Eq. (\ref{interVASR}) for the scenario involving $V-A$ and $S_R$ couplings.  
Second,  even if the differential
cross section is integrated over $\phi_{e}$, but there is  PET,  the energy spectrum of recoil electrons and the distribution of outgoing electrons  polar angle are still  sensitive to the interferences, proportional to the angular correlations among $ \hat{\bf q}, \mbox{\boldmath $(\hat{\eta}_{\nu})^{ \perp}$},  
\mbox{\boldmath $(\hat{\eta}_{e})^{\perp}$}$ vectors, Figs. 7, 8, 9, 10.   
It is worth pointing out that the measurements of the azimuthal asymmetry and of the polar angle distribution require the intense low-energy $\nu_{e}$ sources, the polarized target-electrons, and the detectors  observing both the azimuthal angle and polar angle of the scattered electrons with the good angular resolution. 
The detectors with the very low threshold for the precise measurements of outgoing electrons spectrum would be needed. Let us remind that  the ideas of proper detectors such as Hellaz \cite{Hellaz,Hellaz1,Hellaz2} and Heron \cite{Heron,Heron1,Heron2}  have been considered in the literature. The silicon cryogenic detectors (Neganov et al., hep-ex/0105083),  the high purity germanium detectors, the semiconductor detectors  and the bolometers \cite{semi, bolo} seem to be also  interesting proposals. Our investigation is made in hope to encourage the neutrino laboratories working with the artificial (un)polarized $\nu$ sources, and to revive the discussion on the feasibility of  PET and on the development of ultra-low threshold high-precision detection techniques in the context of TRSV in the low energy leptonic and semileptonic weak interaction processes. 
 
 \section{Appendix 1- General formula on laboratory differential cross section for NEES in PET case}
 
 The formula on the laboratory differential cross section  
calculated with the amplitude $M^{D}_{\nu_{e} e^{-}}$,  Eq. (\ref{ampD}): 
 \beq \label{przekDnue} 
\lefteqn{ \frac{d^{2} \sigma}{d y d \phi_{e}} = \bigg(\frac{d^{2} \sigma}{d y d \phi_{e}}\bigg)_{V- A} + \bigg(\frac{d^{2} \sigma}{d y d \phi_{e}}\bigg)_{V+A} }\\ 
&&\mbox{}  + \bigg(\frac{d^{2} \sigma}{d y d \phi_{e}}\bigg)_{(S,T,P)_R} 
 + \bigg( \frac{d^{2} \sigma}{d y d \phi_{e}} \bigg)_{V-A}^{S_{R}} \nonumber  \\
&&\mbox{} + \bigg( \frac{d^{2} \sigma}{d y d \phi_{e}} \bigg)_{V-A}^{P_{R}} + \bigg( \frac{d^{2} \sigma}{d y d \phi_{e}} \bigg)_{V-A}^{T_{R}} + \bigg( \frac{d^{2} \sigma}{d y d \phi_{e}} \bigg)_{V+A}^{S_{R}} \nonumber\\
&&\mbox{} + \bigg(\frac{d^{2} \sigma}{d y d \phi_{e}} \bigg)_{V+A}^{P_{R}} + \bigg(\frac{d^{2} \sigma}{d y d \phi_{e}} \bigg)_{V+A}^{T_{R}},\nonumber 
\eeq 
\beq \lefteqn{ \bigg( \frac{d^{2} \sigma}{d y d \phi_{e}} \bigg)_{V- A} = 
B (1-\mbox{\boldmath $\hat{\eta}_{\nu}$}\cdot\hat{\bf q}) \bigg\{ 
(c_{A}^{L})^2 \bigg[ (y-2)y + 2}\nonumber\\
&& \mbox{} +   \frac{m_e}{E_\nu} y  -\netpet  \sqrt{ \frac{2 m_{e}}{E_{\nu}}+y} 
\left(\sqrt{y^{3}}-2\sqrt{y}\right)  \bigg]\\
&& \mbox{} - (c_{V}^{L})^2 \bigg[ \netpet\sqrt{y^{3}}\sqrt{\frac{2m_e}{E_\nu}+y}
 - y^{2} \nonumber \\
&& \mbox{} + y\left(\frac{m_e}{E_\nu} + 2\right) -2 \bigg] + 2(c_{V}^L c_A^{L}) \bigg[  (2-y)y \nonumber\\ 
 && \mbox{} + \netpet(y-1)\sqrt{y \left(\frac{2m_e}{E_\nu}+y\right)} \bigg]
\bigg\},\nonumber
\eeq
\beq\lefteqn{\bigg(\frac{d^{2} \sigma}{d y d \phi_{e}}\bigg)_{V+A} = B (1+\mbox{\boldmath $\hat{\eta}_{\nu}$}\cdot\hat{\bf q}) \bigg\{ 
(c_{A}^{R})^2 \bigg[ (y-2)y + 2}\nonumber\\
&& \mbox{} +   \frac{m_e}{E_\nu} y  +\netpet \sqrt{ \frac{2 m_{e}}{E_{\nu}}+y} 
\left(\sqrt{y^{3}}-2\sqrt{y}\right)  \bigg]\\
&& \mbox{} + (c_{V}^{R})^2 \bigg[ \netpet\sqrt{y^{3}}\sqrt{\frac{2m_e}{E_\nu}+y}
 + y^{2} \nonumber \\
&& \mbox{} - y\left(\frac{m_e}{E_\nu} + 2\right) +2 \bigg] + 2(c_{V}^R c_A^{R}) \bigg[  (2-y)y \nonumber\\ 
 && \mbox{} + \netpet(1-y)\sqrt{y \left(\frac{2m_e}{E_\nu}+y\right)} \bigg]
\bigg\},\nonumber\eeq
 \beq
\lefteqn{\bigg(\frac{d^{2} \sigma}{d y d \phi_{e}}\bigg)_{(S,T,P)_{R}} = \mbox{}
B\bigg\{ y\left(y+2\frac{m_{e}}{E_{\nu}}\right)
  |c_{S}^{R}|^{2} + y^{2}|c_{P}^{R}|^{2} }   \\
&& \mbox{} + 2 \bigg[ \bigg( (2-y)^2 -\frac{m_{e}}{E_{\nu}}y 
- \netpet\mbox{\boldmath $\hat{\eta}_{\nu}$}\cdot\hat{\bf q}  \sqrt{ \frac{2 m_{e}}{E_{\nu}}+y} \nonumber\\
&& \mbox{} \cdot \left(\sqrt{y^{3}}-2\sqrt{y}\right) \bigg)|{c_{T}^{R}}|^{2} 
+ y\bigg( (y-2) \nonumber \\ 
&& \mbox{} -\netpet \mbox{\boldmath $\hat{\eta}_{\nu}$}\cdot\hat{\bf q}  
\sqrt{\left(\frac{2 m_{e}}{E_{\nu}}+y\right)y} \bigg) Re(c_{P}^{R} c_{T}^{*R})  \nonumber  \\
&& \mbox{} -  \netpet \mbox{\boldmath $\hat{\eta}_{\nu}$}\cdot\hat{\bf q}  \sqrt{\frac{2 m_{e}}{E_{\nu}}+y}\,\sqrt{y^{3}}\, Re(c_{S}^{R} c_{P}^{*R})\nonumber  \\
&& \mbox{} + (y-2)\bigg( y - \netpet \mbox{\boldmath $\hat{\eta}_{\nu}$}\cdot\hat{\bf q}  \sqrt{y\left(\frac{2 m_{e}}{E_{\nu}}+y\right)} \bigg)  \nonumber \\
&& \mbox{} \cdot  Re(c_{S}^{R}c_{T}^{*R})
 + 2 \sqrt{ y\left(y+2\frac{m_{e}}{E_{\nu}}\right)} \netqpet \nonumber \\
 && \mbox{} \cdot Im(c_{S}^{R}c_{T}^{*R})\bigg] 
\bigg\},  \nonumber\eeq
\beq
 \lefteqn{\label{VmASR} \bigg( \frac{d^{2} \sigma}{d y d \phi_{e}} \bigg)_{V-A}^{S_R} = \mbox{}
B \Bigg\{ 
c_{A}^{L} \frac{E_{\nu}}{m_e}\Bigg[ \sqrt{y\left(\frac{2 m_{e}}{E_{\nu}}+y\right)^{3}} }\\
&& \mbox{}  \cdot\left(\petnetnut Im (c_{S}^{R})\right) -\bigg(y^{2} + \frac{2 m_{e}y}{E_{\nu}}\bigg) \nonumber\\
&& \mbox{}  \cdot\Bigg(\netpet\bigg( \nutpet Re(c_{S}^{R})\nonumber\\
&&\mbox{} -\nutqpet Im (c_{S}^{R}\bigg)  + \qnetnut  \nonumber\\
&&\mbox{} \cdot Im (c_{S}^{R})+ \frac{ m_{e}}{E_{\nu}}\netnut Re(c_{S}^{R}))\Bigg)\Bigg] \nonumber\\
&& \mbox{} +  c_{V}^{L}\Bigg[ \sqrt{\frac{2 m_{e}}{E_{\nu}}+y)y} \Bigg( 2\bigg( \nutpet Re(c_{S}^{R})\nonumber \\
&&\mbox{} - \nutqpet Im (c_{S}^{R})\bigg) \nonumber\\
&& \mbox{} +  y \petnetnut Im (c_{S}^{R})\Bigg) \nonumber\\
&& \mbox{} + y(y -2) \netnut Re(c_{S}^{R})\nonumber\\
&& \mbox{} + \bigg(\frac{E_{\nu}}{m_{e}}y^{2} + 2 y \bigg) \nutpet\bigg(\netpet Re(c_{S}^{R})\nonumber\\
&&\mbox{} - \netqpet Im (c_{S}^{R})\bigg)
\Bigg]
\Bigg\},
\nonumber \eeq
\beq
 \lefteqn{\label{VmAPR} \bigg( \frac{d^{2} \sigma}{d y d \phi_{e}} \bigg)_{V-A}^{P_R} = \mbox{}
B \Bigg\{ 
c_{V}^{L} \frac{E_{\nu}}{m_e}\Bigg[  y \sqrt{y\left(\frac{2 m_{e}}{E_{\nu}}+y\right)} }\\
&& \mbox{}   \cdot\left(\petnetnut Im (c_{P}^{R})\right) + \bigg(y^{2} + \frac{2 m_{e}y}{E_{\nu}}\bigg) \nonumber\\
&& \mbox{}  \cdot\Bigg(\netpet\bigg( - \nutpet Re(c_{P}^{R})\nonumber\\
&&\mbox{} + \nutqpet Im (c_{P}^{R}\bigg)\Bigg)  -  \bigg(\qnetnut  \nonumber\\
&&\mbox{} \cdot Im (c_{P}^{R})+ \frac{ m_{e}}{E_{\nu}}\netnut Re(c_{P}^{R})\bigg)y^2\Bigg] \nonumber\\
&& \mbox{} +  c_{A}^{L}\Bigg[ y 
\sqrt{\left(\frac{2 m_{e}}{E_{\nu}}+y\right)y} \petnetnut  \nonumber\\
&& \mbox{} \cdot Im (c_{P}^{R})+ y(y -2) \netnut Re(c_{P}^{R})\nonumber\\
&& \mbox{} + \bigg(\frac{E_{\nu}}{m_{e}}y^{2} + 2 y \bigg) \nutpet\bigg(\netpet Re(c_{P}^{R})\nonumber\\
&&\mbox{} - \netqpet Im (c_{P}^{R})\bigg)
\Bigg]
\Bigg\},
\nonumber \eeq
\beq
 \lefteqn{\label{VmATR} \bigg( \frac{d^{2} \sigma}{d y d \phi_{e}} \bigg)_{V-A}^{T_R} = \mbox{}
2 B \Bigg\{ 
\frac{E_{\nu}}{m_e}\Bigg[c_{A}^{L}\Bigg(  \sqrt{y\left(\frac{2 m_{e}}{E_{\nu}}+y\right)} }\\
&& \mbox{}  \cdot\left(1-2y + \frac{2 m_{e}y}{E_{\nu}}\right)\petnetnut Im (c_{T}^{R})\nonumber\\
&& \mbox{} + 2 \bigg(y^{2} + \frac{2 m_{e}y}{E_{\nu}}\bigg) 
 \bigg(\nutpet \netqpet   \nonumber\\
&& \mbox{}  - \netpet \nutqpet\bigg) Im(c_{T}^{R})\nonumber\\
&&\mbox{}  + (2y-1)\left( y +\frac{2 m_{e}}{E_{\nu}}\right)\qnetnut Im (c_{T}^{R}) \nonumber\\
&&\mbox{} - \left(\frac{ m_{e}}{E_{\nu}}\right)(y-2)\netnut Re(c_{T}^{R})\nonumber\\
 && \mbox{} + \left(\frac{ m_{e}}{E_{\nu}}\right)\sqrt{y\left(\frac{2 m_{e}}{E_{\nu}}+y\right)} \bigg(\nutpet Re(c_{T}^{R})\nonumber\\
 && \mbox{} - \nutqpet Im(c_{T}^{R})\bigg) \Bigg) \nonumber\\
&& \mbox{} +  c_{V}^{L} \Bigg(\left( 2(1-y) + \left(\frac{ m_{e}}{E_{\nu}}\right)(1-2y)\right)\sqrt{\left(\frac{2 m_{e}}{E_{\nu}}+y\right)y} \nonumber\\ 
&& \mbox{} \cdot \left(\petnetnut Im (c_{T}^{R})\right) +  2\bigg( y^{2} +  \frac{2 m_{e}y}{E_{\nu}}\bigg) \nonumber \\ 
&& \mbox{} \cdot  \bigg(\nutpet \netqpet \nonumber\\
&& \mbox{} - \netpet\nutqpet \bigg)Im (c_{T}^{R}) \nonumber\\
&&\mbox{} +   2\left( y-1\right) \left( y+\frac{ m_{e}}{E_{\nu}}\right)\qnetnut Im (c_{T}^{R})\Bigg] \nonumber\\
&& \mbox{} + \left(2-y +\frac{ m_{e}y}{E_{\nu}}\right)  \netnut Re(c_{T}^{R})\Bigg)\Bigg\},\nonumber
\eeq
\beq
 \lefteqn{\label{VpASR} \bigg( \frac{d^{2} \sigma}{d y d \phi_{e}} \bigg)_{V+A}^{S_R} = \mbox{}
B \Bigg\{ 
c_{A}^{R} \frac{E_{\nu}}{m_e}\Bigg[ \sqrt{y\left(\frac{2 m_{e}}{E_{\nu}}+y\right)^{3}} }\\
&& \mbox{}  \cdot\left(\petnetnut Im (c_{S}^{R})\right) -\bigg(y^{2} + \frac{2 m_{e}y}{E_{\nu}}\bigg) \nonumber\\
&& \mbox{}  \cdot\Bigg(\netpet\bigg( \nutpet Re(c_{S}^{R})\nonumber\\
&&\mbox{} -\nutqpet Im (c_{S}^{R}\bigg)  + \qnetnut  \nonumber\\
&&\mbox{} \cdot Im (c_{S}^{R})+ \frac{ m_{e}}{E_{\nu}}\netnut Re(c_{S}^{R}))\Bigg)\Bigg] \nonumber\\
&& \mbox{} +  c_{V}^{R}\Bigg[ \sqrt{\left(\frac{2 m_{e}}{E_{\nu}}+y\right)y} \Bigg( 2\bigg( \nutpet Re(c_{S}^{R})\nonumber \\
&&\mbox{} - \nutqpet Im (c_{S}^{R})\bigg) \nonumber\\
&& \mbox{} +  y \petnetnut Im (c_{S}^{R})\Bigg) \nonumber\\
&& \mbox{} + y(y -2) \netnut Re(c_{S}^{R})\nonumber\\
&& \mbox{} + \bigg(\frac{E_{\nu}}{m_{e}}y^{2} + 2 y \bigg) \nutpet\bigg(\netpet Re(c_{S}^{R})\nonumber\\
&&\mbox{} - \netqpet Im (c_{S}^{R})\bigg)
\Bigg]
\Bigg\},
\nonumber  \eeq
\beq
 \lefteqn{\label{VpAPR} \bigg( \frac{d^{2} \sigma}{d y d \phi_{e}} \bigg)_{V+A}^{P_R} = \mbox{}
B \Bigg\{ 
c_{V}^{R} \frac{E_{\nu}}{m_e}\Bigg[  y \sqrt{y\left(\frac{2 m_{e}}{E_{\nu}}+y\right)} }\\
&& \mbox{}   \cdot\left(\petnetnut Im (c_{P}^{R})\right) + \bigg(y^{2} + \frac{2 m_{e}y}{E_{\nu}}\bigg) \nonumber\\
&& \mbox{} \cdot\Bigg( \netpet\bigg( - \nutpet Re(c_{P}^{R})\nonumber\\
&&\mbox{} + \nutqpet Im (c_{P}^{R}\bigg)\Bigg)  -  \bigg(\qnetnut  \nonumber\\
&&\mbox{} \cdot Im (c_{P}^{R})+\frac{ m_{e}}{E_{\nu}}\netnut Re(c_{P}^{R})\bigg)y^2\Bigg] \nonumber\\
&& \mbox{} +  c_{A}^{R}\Bigg[ y 
\sqrt{\left(\frac{2 m_{e}}{E_{\nu}}+y\right)y} \petnetnut  \nonumber\\
&& \mbox{} \cdot Im (c_{P}^{R})+ y(y -2) \netnut Re(c_{P}^{R})\nonumber\\
&& \mbox{} + \bigg(\frac{E_{\nu}}{m_{e}}y^{2} + 2 y \bigg) \nutpet\bigg(\netpet Re(c_{P}^{R})\nonumber\\
&&\mbox{} - \netqpet Im (c_{P}^{R})\bigg)
\Bigg]
\Bigg\},
\nonumber \eeq
\beq
 \lefteqn{\label{VpATR} \bigg( \frac{d^{2} \sigma}{d y d \phi_{e}} \bigg)_{V+A}^{T_R} = \mbox{}
2 B \Bigg\{ 
\frac{E_{\nu}}{m_e}\Bigg[c_{A}^{R} \Bigg( \sqrt{y\left(\frac{2 m_{e}}{E_{\nu}}+y\right)} }\\
&& \mbox{}  \cdot\left(1-2y + \frac{2 m_{e}y}{E_{\nu}}\right)\petnetnut Im (c_{T}^{R})\nonumber\\
&& \mbox{} + 2 \bigg(y^{2} + \frac{2 m_{e}y}{E_{\nu}}\bigg) 
 \bigg(\nutpet \netqpet   \nonumber\\
&& \mbox{}  - \netpet \nutqpet\bigg) Im(c_{T}^{R})\nonumber\\
&&\mbox{}  + (2y-1)\left( y +\frac{2 m_{e}}{E_{\nu}}\right)\qnetnut Im (c_{T}^{R}) \nonumber\\
&&\mbox{} - \left(\frac{ m_{e}}{E_{\nu}}\right)(y-2)\netnut Re(c_{T}^{R})\nonumber\\
 && \mbox{} + \left(\frac{ m_{e}}{E_{\nu}}\right)\sqrt{y\left(\frac{2 m_{e}}{E_{\nu}}+y\right)} \bigg(-\nutpet Re(c_{T}^{R})\nonumber\\
 && \mbox{} + \nutqpet Im(c_{T}^{R})\bigg)\Bigg)  \nonumber\\
&& \mbox{} +  c_{V}^{R} \Bigg(\left( 2(1-y) + \left(\frac{ m_{e}}{E_{\nu}}\right)(1-2y)\right)\sqrt{\left(\frac{2 m_{e}}{E_{\nu}}+y\right)y} \nonumber\\ 
&& \mbox{} \cdot \left(\petnetnut Im (c_{T}^{R})\right) +  2\bigg( y^{2} +  \frac{2 m_{e}y}{E_{\nu}}\bigg) \nonumber \\ 
&& \mbox{} \cdot  \bigg(\nutpet \netqpet \nonumber\\
&& \mbox{} - \netpet\nutqpet \bigg)Im (c_{T}^{R}) \nonumber\\
&&\mbox{} +   2\left( y-1\right) \left( y+\frac{ m_{e}}{E_{\nu}}\right)\qnetnut Im (c_{T}^{R})\Bigg] \nonumber\\
&& \mbox{} + \left(2-y +\frac{ m_{e}y}{E_{\nu}}\right)  \netnut Re(c_{T}^{R})\Bigg)\Bigg\}.\nonumber 
\eeq
\beq y & \equiv &
\frac{T_e}{E_\nu}=\frac{m_{e}}{E_{\nu}}\frac{2 cos^{2}\theta_{e}}
{(1+\frac{m_{e}}{E_{\nu}})^{2}-cos^{2}\theta_{e}} 
\eeq 
is the ratio of the
kinetic energy of the recoil electron $T_{e}$  to the incoming $\nu_e$ 
energy $E_{\nu}$.  
 $m_{e}$ is the
electron mass; $B\equiv \left(E_{\nu}m_{e}/4\pi^2\right) \left(G_{F}^{2}/2\right)$.  $\mbox{\boldmath $\hat{\eta}_{\nu}$}$ is the unit 3-vector of
 $\nu_{e}$ spin  polarization in its rest frame. $(\mbox{\boldmath$\hat{\eta}_{\nu}$}\cdot\hat{\bf q}){\bf\hat{q}}$ is the longitudinal component of $\nu_{e}$ spin polarization.  
$|\mbox{\boldmath $\hat{\eta}_{\nu}$}\cdot\hat{\bf q}| = |1- 2
Q_{L}^{\nu}|$, where $Q_{L}^{\nu}$ is the probability of producing the LC $\nu_{e}$.    

\section{Appendix 2 - Definitions of the asymmetry functions}

The asymmetry function $A(\Phi)$ is defined as
\beq
A(\Phi) := \frac{\int\limits_{\Phi}^{\Phi+\pi}\frac{d\sigma}{d \phi_{e}}\,d\phi_e -
\int\limits_{\Phi+\pi}^{\Phi+2\pi}\frac{d\sigma}{d \phi_{e}}\,d\phi_e}
{\int\limits_{\Phi}^{\Phi+\pi}\frac{d\sigma}{d \phi_{e}}\,d\phi_e +
\int\limits_{\Phi+\pi}^{\Phi+2\pi}\frac{d\sigma}{d \phi_{e}}\,d\phi_e}.
\eeq
Two other asymmetry functions are employed:
\beq
A_y(\Phi) := \frac{\int\limits_{\Phi}^{\Phi+\pi}\frac{d^2\sigma}{d \phi_{e} dy}\,d\phi_e -
\int\limits_{\Phi+\pi}^{\Phi+2\pi}\frac{d^2\sigma}{d \phi_{e} dy}\,d\phi_e}
{\int\limits_{\Phi}^{\Phi+\pi}\frac{d^2\sigma}{d \phi_{e} dy}\,d\phi_e +
\int\limits_{\Phi+\pi}^{\Phi+2\pi}\frac{d^2\sigma}{d \phi_{e} dy}\,d\phi_e},
\eeq
\beq
A_{\theta_e}(\Phi) := \frac{\int\limits_{\Phi}^{\Phi+\pi}\frac{d^2\sigma}{d \phi_{e}d\theta_e}\,d\phi_e -
\int\limits_{\Phi+\pi}^{\Phi+2\pi}\frac{d^2\sigma}{d \phi_{e} d\theta_e}\,d\phi_e}
{\int\limits_{\Phi}^{\Phi+\pi}\frac{d^2\sigma}{d \phi_{e} d\theta_e}\,d\phi_e +
\int\limits_{\Phi+\pi}^{\Phi+2\pi}\frac{d^2\sigma}{d \phi_{e} d\theta_e}\,d\phi_e}.
\eeq


\begin{thebibliography}{99}
\bibitem{SM} S. L. Glashow,  Nucl. Phys.  {\bf 22},  579 (1961)
 \bibitem{SM1} S. Weinberg,  Phys. Rev. Lett.  {\bf 19}, 1264 (1967)
\bibitem{SM2}  A. Salam, A. Salam, in  {\sl Elementary Particle Theory} (Almquist and Wiksells, Stockholm, 1969)
\bibitem{SM3} R. P. Feynman, M. Gell-Mann, Phys. Rev. {\bf 109},    193 (1958)
\bibitem{SM4} E. C. G. Sudarshan, R. E. Marshak, Phys. Rev. {\bf 109},   1860 (1958)
\bibitem{CP} J.H. Christenson, J.W. Cronin, V.L. Fitch and R. Turlay,
Phys. Rev. Lett. {\bf 13},  138 (1964)
\bibitem{CP1} B. Aubert et al., Phys. Rev. Lett. {\bf 87},  091801 (2001) 
\bibitem{CP2} K. Abe et al., Phys. Rev. Lett. {\bf 87}, 091802  (2001)
\bibitem{Kobayashi} M. Kobayashi, T. Maskawa, Prog. Theor. Phys. {\bf 49},  652 (1973)
 \bibitem{barion} A. Riotto, M. Trodden, Annu. Rev. Nucl. Part. Sci. {\bf 49},  35 (1999)
\bibitem{Pati} J.C. Pati, A. Salam, Phys. Rev. D {\bf 10}, 275 (1974)
 \bibitem{Pati1} R. Mohapatra, J.C. Pati, Phys. Rev. D {\bf 11}, 566 (1975); Phys. Rev. D {\bf 11}, 558 (1975) 
\bibitem{Pati2} R.N. Mohapatra, G. Senjanovic, Phys. Rev. D {\bf 12}, 1502 (1975); Phys Rev D {\bf 23}, 165 (1981)
\bibitem{Pati3} M. A. B. Beg et al., Phys. Rev. Lett. {\bf 38}, 1252 (1977)
\bibitem{Pati4} P. Herczeg,  Phys. Rev.  D {\bf 34}, 3449 (1986)
 \bibitem{Jodidio} A. Jodidio et al.,  Phys. Rev.  D {\bf 34},  1967 (1986)
\bibitem{CM} E. J. Eichten, K. D. Lane, M.E. Peskin, Phys. Rev. Lett. {\bf 50}, 811 (1983)
\bibitem{CM1} P. Herczeg, Prog. Part. Nucl. Phys. {\bf 46}, 413 (2001)
  \bibitem{Extra} N. Arkani-Hamed, S. Dimopoulous, G. Dvali, J. March-Russell, Phys. Lett. B {\bf 429}, 263 (1998). 
  \bibitem{unparticle}  T. Banks, A. Zaks, Nucl. Phys. B {\bf 196}, 189 (1982) 
\bibitem{unparticle1} H. Georgi, Phys. Rev. Lett. {\bf 98}, 221601 (2007)
\bibitem{unparticle2}  H. Georgi, Phys. Lett. B {\bf 650}, 275 (2007)
\bibitem{unparticle3} K. Cheung, W.Y. Keung, T.C. Yuan, Phys. Rev. Lett {\bf 99}, 051803 (2007) 
\bibitem{unparticle4} K. Cheung, W.Y. Keung, T.C.
Yuan, Phys. Rev. D {\bf 76}, 055003 (2007) 
\bibitem{unparticle5} S. L. Chen, X. G. He, Phys. Rev.  D {\bf 76}, 091702 (2007)
\bibitem{unparticle6} A. B. Balantekin, K. O. Ozansoy, Phys. Rev. D {\bf 76}, 095014 (2007) 
\bibitem{unparticle7} J. Barranco et al., Phys. Rev.  D {\bf 79}, 073011 (2009) 
\bibitem{unparticle8} D. Montanino, M. Picariello, J. Pulido, Phys. Rev. D {\bf 77}, 093011 (2008)
\bibitem{unparticle9} S. Zhou, Phys. Lett. B {\bf 659}, 336 (2008)
\bibitem{unparticle10} B. Grinstein, K. A. Intriligator,  I. Z. Rothstein, Phys. Lett. B {\bf 662}, 367 (2008)
\bibitem{unparticle11} M. Deniz et al., Phys. Rev.  D {\bf 82}, 033004 (2010)
\bibitem{unparticle12} J. Barranco et al., Int. J. Mod. Phys. A {\bf 27}, 1250147 (2012)
\bibitem{Geer} S. Geer, Phys. Rev. D {\bf 57}, 6989 (1998) 
\bibitem{Geer1} S. Geer, Phys. Rev. D {\bf 59},  039903 (1999)
\bibitem{neutron} L. J. Lising et al.,  Phys. Rev.  C  {\bf 62},  055501 (2000)
\bibitem{Herczeg} P. Herczeg, J. Res. Nath. Inst. Stand. Technol. {\bf 110},  453 (2005)
\bibitem{Huber} R. Huber et al., Phys. Rev. Lett. {\bf 90}, 202301 (2003)
\bibitem{Bodek} K. Bodek et al., J. Res. Natl. Inst. Stand. Technol. {\bf 110},  461 (2005) 
\bibitem{Mumm} H. P. Mumm et al., Phys. Rev. Lett. {\bf 107},  102301 (2011)
\bibitem{NDM} S. Weinberg, Phys. Rev. D {\bf 42},  860 (1990)
 \bibitem{NDM1} R. Garisto, G. L. Kane, Phys. Rev. D {\bf 44},  2038 (1991)
 \bibitem{NDM2} G. Belanger, C. Q. Geng, Phys. Rev. D {\bf 44},  2789 (1991)
 \bibitem{NDM3} G. H. Wu, J. N. Ng. Phys.Lett. B {\bf 392}, 93 (1997) 
\bibitem{NDM4} E. Gabrielli, Phys. Lett. B {\bf 301}, 409 (1993) 
   \bibitem{INFN} B. Babussinov et al.,  Nucl. Instrum. and Meth. A {\bf 694}, 335 (2012) 
\bibitem{Misiaszek} M. Misiaszek et al., Nucl. Phys.  B {\bf 734}, 203 (2006) 
\bibitem{PET} J. Bernabeu et al., Phys. Lett.  B {\bf 613}, 162 (2005) 
 \bibitem{PET1} V. A. Guseinov et al., Phys. Rev.  D {\bf 75}, 073021 (2007)
\bibitem{PET2} S. Ciechanowicz et al., Phys. Rev.  D {\bf 71},  093006 (2005) 
\bibitem{PET3}P. Minkowski, M. Passera, Phys. Lett.  B {\bf 541},  151 (2002) 
\bibitem{PET4} T. I. Rashba, V. B. Semikoz,  Phys. Lett.  B {\bf 479},  218 (2000) 
\bibitem{PET5} W.-T. Ni et al., Phys. Rev. Lett. {\bf 82},  2439 (1999)
 \bibitem{PET6} W. Bialek et al., Phys. Rev. Lett. {\bf 56}, 1623 (1986)  
\bibitem{PET7} P. V. Vorobyov, Y. I. Gitarts, Phys. Lett.  B {\bf 208}, 146 (1988) 
\bibitem{Data} K.A. Olive et al. (Particle Data Group), Chin. Phys. C {\bf 38}, 090001 (2014) 
\bibitem{SoxBell} G. Bellini et al., JHEP {\bf 08}, 038 (2013) 
\bibitem{sterile} C. Athanassopoulos et al., Phys. Rev. Lett. {\bf 75}, 2650 (1995)
\bibitem{sterile1} C. Athanassopoulos et al., Phys. Rev. C {\bf 54}, 2685 (1996)
\bibitem{sterile2} A. Aguilar et al., Phys. Rev. D {\bf 64}, 112007 (2001) 
\bibitem{sterile3}  A. Anguilar-Arevalo et al., Phys. Rev. Lett. {\bf 105}, 181801 (2010)
\bibitem{sterile4} Th. Mueller et al., Phys. Rev. C {\bf 83}, 054615 (2011) 
\bibitem{sterile5} P. Huber, Phys. Rev. C {\bf 84}, 024617 (2011)
 \bibitem{sterile6} G. Mention et al., Phys. Rev. D {\bf 83}, 073006 (2011)
\bibitem{CMS2003} S. Ciechanowicz, M. Misiaszek, S. Sobk\'ow, Eur. Phys. J. C {\bf 32}, s01, s151 (2003)
 \bibitem{Mulan} D. M. Webber et al.,  Phys. Rev. Lett. {\bf 106}, 041803 (2011)
\bibitem{Michel} L. Michel, A. S. Wightman,   Phys. Rev. {\bf 98},  1190 (1955) 
 \bibitem{Hellaz} F. Arzarello et al., Report No. CERN-LAA/94-19, College de France LPC/94-28, 1994
\bibitem{Hellaz1} J. Seguinot et al., Report No. LPC 95 08, College de France, Laboratoire de
Physique Corpusculaire, 1995 
\bibitem{Hellaz2} A. Sarrat, Nucl. Phys. Proc. Suppl. {\bf 95}, 177 (2001)
\bibitem{Heron} R. E. Lanou et al., The Heron project. Abstracts of Papers of the American Chemical
Society 2(217), 021-NUCL 1999
\bibitem{Heron1} Y.H. Huang, R. E. Lanou, H. J. Maris, G. M. Seidel, B. Sethumadhavan,
 W. Yao,  Astropart. Phys. {\bf 30}, 1 (2008)
\bibitem{Heron2} J. S. Adams, Y. H. Huang, Y. H. Kim, R. E. Lanou, H. J. Maris, 
G. M. Seidel, The HERON project, chapter 8, pages 70-80, 2002 
 \bibitem{semi} C. E. Aalseth et al., Phys. Rev. Lett. {\bf 106}, 131301 (2011)
\bibitem{bolo} C. Enss, {\sl Cryogenic Particle Detection} (Springer-Verlag Berlin
Heidelberg, 2005)
  

\end{thebibliography}
\end{document}